# Temporal Series Analysis Approach to Spectra of Complex Networks

Huijie Yang[*], Fangcui Zhao, Longyu Qi, Beilai Hu

School of Physics, Nankai University, Tianjin 300071, China

**Abstract**

The spacing of nearest levels of the spectrum of a complex network can be regarded as a time series. Joint use of Multi-fractal Detrended Fluctuation Approach (MF-DFA) and Diffusion Entropy (DE) is employed to extract characteristics from this time series. For the WS (Watts and Strogatz) small-world model, there exist a critical point at rewiring probability $p_r = 0.32$. For a network generated in the range $0 \prec p_r \prec 0.32$, the correlation exponent is in the range of $1.0 \sim 1.64$. Above this critical point, all the networks behave similar with that at $p_r = 1$. For the ER model, the time series behaves like FBM (fractional Brownian motion) noise at $p_{ER} = 1/N$. For the GRN (growing random network) model, the values of the long-range correlation exponent are in the range of $0.74 \sim 0.83$. For most of the GRN networks the PDF of a constructed time series obeys a Gaussian form. In the joint use of MF-DFA and DE, the shuffling procedure in DE is essential to obtain a reliable result.

PACS number(s): 89.75.-k , 05.45.-a, 02.60.-x

---

[*] Corresponding author, E-mail address: huijieyangn@eyou.com

# I. INTRODUCTION

Detailed investigations indicate that real-world networks have highly distinctive statistical signatures very far from random network [1]. Two classes of models, called the small-world graphs and the scale-free networks, are proposed to capture the clustering and the power-law degree distribution present in many real networks, respectively [2-5]. However, most analyses have been confined to capture the static structural properties, e.g., degree distribution, shortest connecting paths, clustering coefficients, etc. Capturing the global characteristics of complex networks is an essential role at present time. Another problem is the lack of suitable techniques, which leaves a large gap in our capturing the basic properties comprehensively and understanding networks theoretically. Thus, another important role is to use concepts or techniques developed in other fields to characterize complex networks.

It is demonstrated in extensive literature that the properties of graphs and the associated adjacency matrices are well characterized by spectral methods. Investigations on spectrum can provide global measures of the network properties [6-16]. Actually, analyzing spectrum is one of the most important tools to understand comprehensively the dynamical processes in a complex quantum-mechanical system [17-21]. In recent literature, it is pointed out that joint use of variance-based detectors and the DE (diffusion entropy) analysis is a powerful tool to capture the scaling invariance embedded in a time series [22]. In this paper, regarding the spacing of nearest levels of a spectrum as a time series, we try to detect the self-similar structures and long-range correlations embedded in the spectrum of the adjacency matrices of complex networks by means of joint use of DE and MF-DFA (multifractal detrended fluctuation approach).

# II. METHODS

A complex network $G$ can be represented by its adjacency matrix $A(G)$. For an undirected complex network $A(G)$ should be a real symmetric matrix: $A_{ij} = A_{ji} = 1$, if nodes $i$ and $j$ are connected, or $0$, if these two nodes are not connected. The main algebraic tool that we will use for the analysis of complex networks will be the spectrum, i.e., the set of eigenvales of the complex network's adjacency matrix, called the spectrum of the complex network. Denoting this spectrum as $\{E_0, E_1, E_2, \ldots E_N\}$, we can construct a time series with the intervals between two successive eigenvales as,

$$\{\Delta E_k | k = 1, 2, \cdots N+1\} = \{E_1 - E_0, E_2 - E_1, \cdots E_N - E_{N-1}, E_0 - E_N\} \quad (1)$$

The MF-DFA method [23-25] is used to measure the long-range correlation. The origin spectrum $\{E_0, E_1, E_2, \ldots E_N\}$ can be employed as the profile of the constructed time series. Connecting the starting and the end of this profile, we can obtain all possible segments with length $l$, $\{(E_m, E_{m+1}, \ldots E_{m+l-1}) | m = 0, 1, 2, \cdots N\}$. Fit each segment with a $r$-order polynomial function. The fitting result can be regarded as the local trends of all the segments. Taking the local trends out from the corresponding segments, if there exist long-range correlation the variance will obey a power-law, that is,

$$v(l, r, q,) = \frac{1}{2(N+1)} \cdot \sum_{m=0}^{N} \left[ \frac{1}{l} \cdot \sum_{s=1}^{l} (E_{m+s-1} - E_F^m(s))^2 \right]^{q/2}$$

$$V(l, r, q) = v(l, r, q)^{1/q} \propto l^{\alpha(q, r)} \quad (2)$$

Where $E_F^m$ is the fitting result for the $m$'th segment. If $\alpha(2, r) = 0.5$, there is no correlation and the signal is an uncorrelated signal (white noise); if $\alpha(2, r) \prec 0.5$, the signal is anti-correlated; if $\alpha(2, r) > 0.5$, there is a positive correlation in the signal If the analyzed signal behaves like Brownian noise, we have $\alpha(2, r) = 1.5$. It should be noted that overlapping windows are used in this paper instead of the non-overlapping procedure in dividing the profile into segments [26].

The concept of DE [22,27-29] is also used to find self-similar structures. Connecting the starting and the end of the initially constructed time series, we can obtain a set of delay register vectors as,

$$\{E_1 - E_0, E_2 - E_1, \ldots E_n - E_{n-1}\}$$

$$\{E_2 - E_1, E_3 - E_2, \ldots E_{n+1} - E_n\}$$

$$\vdots$$

$$\{E_0 - E_N, E_1 - E_0, \ldots E_{n-1} - E_{n-2}\} \quad (3)$$

Considering each vector as a trajectory of a particle in duration of $n$ time units, all the above vectors can be regarded as a diffusion process for a system with $N+1$ particles. Accordingly, for each time denoted with $n$ we can reckon the distribution of the displacements of all the particles as the state of the system at time $n$. Dividing the possible range of displacements into $M_0$ bins, DE approach defines diffusion entropy as,

$$S(n) = -\sum_{m=1}^{M_0} \frac{K_m(n)}{N+1} \ln\left(\frac{K_m(n)}{N+1}\right). \tag{4}$$

Where $K_m(n)|m=1,2,...M_0$ is the number of particles whose displacements fall in the $m$'th bin at time $n$. Assume the probability distribution function (PDF) of this diffusion process fulfills the scaling property,

$$p(m,n) = \frac{K_m(n)}{N+1} = \frac{1}{n^\delta} F\left(\frac{m}{n^\delta}\right) | m = 1,2,...M_0. \tag{5}$$

Change the sum operation to integration. After some trivial change of integration we get,

$$S(n) = A + \delta \cdot \ln n, \tag{6}$$

Where $A$ is a constant depending on the function form of PDF. To obtain a suitable $M_0$, the size of a cell is chosen to be a fraction of the square root of the variance of the constructed time series, which reads, $\varepsilon = \sqrt{\frac{\sum_{k=1}^{N+1} (\Delta E_k)^2}{N+1}}$.

In DE approach the method adopted to define the trajectories is based on the idea of a moving window of size $n$ that makes the $s$'th trajectory closely correlated to the next, the $(s+1)$'th trajectory. The two trajectories have $n-1$ values in common. Just as pointed out in the designer's works, the motivation for using overlapping windows is given by their wish to establish a connection with the Kolmogorov-Sinai (KS) entropy [30-31]. Moving a window of size $n$ along a symbolic sequence, we can construct all the possible combination of symbols, and from the frequency of each combination it is possible to derive the Shannon entropy $SE(n)$. The KS entropy can be obtained by the asymptotic limit $\lim_{n \to \infty} SE(n)/n$. It is believed that the same sequence, analyzed with the DE method, at the large values of $n$ where finite KS entropy shows up, must yield a well-defined scaling $\delta$.

Because of the periodic condition the displacements at time $n$ can be written as,

$$D_s(n) = \sum_{i=s}^{s+n-1}(E_i - E_{i-1}) \Big| s = 1,2,3,...N+1. \tag{7}$$

On the other hand, the displacements at time $N-n$ are,

$$\begin{aligned}
D_s(N-n) &= \sum_{i=s}^{s+N-n-1}(E_i - E_{i-1}) \Big| s = 1,2,3,...,N+1 \\
&= \sum_{i=s}^{s+N-1}(E_i - E_{i-1}) - \sum_{i=s+N-n}^{s+N-1}(E_i - E_{i-1}) \Big| s = 1,2,3,...,N+1 \\
&= 0 - \sum_{i=s+N-n}^{s+N-1}(E_i - E_{i-1}) \Big| s = 1,2,3,...,N+1 \\
&= -D_{s+N-n}(n) \Big| s = 1,2,3,...,N+1
\end{aligned} \tag{8}$$

Hence the shape of PDF at time $n$ is identical with that at time $N-n$, the DE results are symmetric with respect to the time point $n = N/2$.

The DE approach can give a right result only when the time series is stationary. Shuffling the time series can eliminate the effects of non-stationary and other kind of correlations among the elements.

The DE can capture exactly the real scaling exponent $\delta$ in Eq.(5) for any function form of PDF. But the MF-DFA can capture it only for some special forms of PDF, such as Gaussian. That is, generally, $\delta \neq \alpha(2,r)$. For Gaussian form, $\delta = \alpha(q,r)$. For Levy walk process, $\delta = \dfrac{1}{3 - 2\alpha(2,r)}$. Joint use of DE and MF-DFA will reveal important information about the PDF.

The adjacency matrices are diagonalized with the Matlab version of the software package PROPACK [32].

### III. RESULTS

**A. Erdos-Renyi model**

Consider the Erdos-Renyi model [33]. Starting with $N$ nodes and no edges, connect each pair with probability $p_{ER}$. It is demonstrated that there exist a critical point $(p_c = 1/N)$ for this kind of random networks. For $p_{ER} \prec p_c$ the network is broken into many small clusters, while at $p_c$ a large cluster forms, which in the asymptotic limit contains all nodes [34].

For $p_{ER} \prec p_c$, the adjacency matrix of the Erdos-Renyi network can be reduced into many small sub-matrices. There is not long-range correlation in its corresponding spectrum. For $p_{ER} \succ p_c$, almost all the nodes belong to one cluster and the connectivity probability for each pair of nodes is $p_{ER}$. The ER network with $p_{ER} = \frac{2j}{N} (j \geq 1)$ is equivalent with a complete random network constructed with WS small-world model ($p_r = 1, k = j$). Hence, we can predict the exponents as,

$$(\alpha(2,2), \delta_{shuffling}) = \begin{cases} (0.5, 0.5) & (p_{ER} \prec 1/N) \\ (1.0, 0.5) & (p_{ER} \succ 1/N) \end{cases} \quad (9)$$

Simulation results presented in Fig.(1) and Fig.(2) are consistent with this theoretical prediction. At $p_{ER} = p_c = 1/N$ there exist two scaling regimes in MF-DFA results. The scaling exponents in the short and long regimes are 0.81 and 0.64, respectively. The corresponding value of shuffling DE is 0.66. In the long regime we have $\alpha(2,2) = \delta$. The time series constructed should behave like FBM (fractional Brownian Motion) noise.

**B. WS Small-world model**

Consider the small-world model introduced by Watts and Strogatz (WS) [1-5]. Adopt the one-dimensional lattice model of the small-world network. That is, take a one-dimensional lattice of $L$ nodes with periodic boundary conditions, and join each node with its $k$ right-handed nearest neighbors. Going through each edge in turn and with probability $p_r$ rewiring one end of this edge to a new node chosen randomly. During the rewiring procedure double edges and self-edges are forbidden. Numerical simulations by Watts and Strogatz show that this rewiring process allows the small-world model to interpolate between a regular lattice and a random graph with the constraint that the minimum degree of each node is fixed [2]. The parameter $k$ is chosen to be 2, and $L$ 3000.

Fig.(3) shows several typical MF-DFA results for different values of $p_r$, i.e., $p_r = 0.0, 0.2, 0.3, 0.8$. Fig.(4) presents the values of scaling exponent $\alpha(2,2)$ for $p_r \in [0,1]$.

For $p_r = 0$ the generated network is regular and periodical, we have $\alpha(2,2) = 1.64$, the time series can be regarded as a slight deviation from Brownian noise. In the range $0.32 \leq p_r \leq 1.0$, $\alpha(2,2) \approx 1 \pm 0.1$. In the range $0 \prec p_r \prec 0.32$, two scaling regimes can be found. That is, there exist a transition point at $p_r = 0.32$. Denoting the values of the scaling exponents in the short and long regimes with $\alpha_1(2,2)$ and $\alpha_2(2,2)$, we have,

$$0.78 \prec \alpha_1(2,2) \leq 1.1,$$

$$1.0 \prec \alpha_2(2,2) \leq 1.64. \qquad (10)$$

Table (1) presents the shuffling DE, un-shuffling DE and SDA results in detail for different values of $p_r$. Shuffling DE show that the values of the scaling exponent $\delta$ are in the range of $[0.46, 0.58]$. Joint use of MF-DFA and shuffling DE tells us that the corresponding PDF obeys a scaling invariant form rather than a Gaussian or a Levy walk one. The scaling exponents derived from un-shuffling DE are all significantly larger than the corresponding shuffling ones, which may be induced by the long-range correlations between elements. Clearly, joint use of the SDA and the un-shuffling DE results cannot give a reliable PDF form.

### C. Growing random network model

Consider the growing random network (GRN) model [4,35]. At each time step, a new site is added and a link to one of the earlier sites is created. The connection kernel $A_k$, defined as the probability that a newly introduced site links to a pre-existing site with $k$ links, determines the structure of this graph. Consider the complex networks generated with a class of homogeneous connection kernels, $A_k \propto k^\theta (0 \leq \theta \leq 1)$. The connectivity distribution decreases as a stretched exponential in $k$, and the asymptotic behavior of which shows two critical points at $\theta_1 = \frac{1}{3}$ and $\theta_1 = \frac{1}{2}$.

Networks with different $\theta$ are generated. The size of each network is selected to be 4000.

It is found that there exist long-range correlation effects in all these constructed time series from $\theta = 0$ to $1$. The power-law is obeyed almost exactly. Fig.(5) shows several typical MF-DFA results for different values of $\theta$, i.e., $\theta = 0.0, 0.2, 0.4, 0.8$.

Fig.(6) presents several typical DE results for different values of $\theta$, i.e., $\theta = 0.0, 0.2, 0.4, 0.8$. To obtain reliable result, we shuffle the constructed time series firstly to eliminate the effects of correlations between elements. For most of the generated networks, shuffling DE approach can detect two scaling regimes. Denote the scaling exponents for short and long regimes with $\delta_1$ and $\delta_2$. Basically, we have $\delta_1 \succ \delta_2$, significantly.

To check the effect of correlations between elements, un-shuffling result is also presented. This effect is such strong that the shuffling procedure is essential to obtain a reliable result. The un-shuffling result cannot distinguish the two scaling regions.

Table (2) presents the DE, SDA and MF-DFA results in detail for different values of $\theta$. Comparison between the shuffling DE (the short regime $\delta_1$) and the MF-DFA results shows that for most values of $\theta$ the PDF has a Gaussian form. At each of the points $\theta = 0.0, 0.15, 0.20, 0.5, 0.55, 0.60$ the PDF has a Levy walk form. There are also several points at which we cannot find a preferred PDF form at present time.

## IV. CONCLUSIONS

For WS model, the long-range correlation exponents for the two limit conditions, the regular network ($p_r = 0$) and complete random network ($p_r = 1$), are $1.64$ and $1.0$, respectively. For a network generated in the range $0 \prec p_r \prec 0.32$, the correlation exponent is in the range of $1.0 \sim 1.64$. Above the critical point $p_r = 0.32$, all the networks behave similar with the complete random one ($p_r = 1$). Joint use of shuffling DE and MF-DFA cannot determine the PDF of the constructed time series.

For ER model, a network with $p_{ER} \succ 1/N$ is similar with a complete random WS network. At the critical point $p_{ER} = 1/N$, the constructed time series behaves like FBM (fractional Brownian motion) noise.

For GRN model the values of the long-range correlation exponent are in the range of $0.74 \sim 0.83$. Joint use of MF-DFA and shuffling DE can give the PDF form in most cases. The average connectivity probability is $2/N$, but the connection kernel determines the connectivity probability for each pair of nodes. Because of the growing character of the GRN model, we cannot simply regard a GRN network with $\theta = 0$ as an ER network at $p_{ER} = 2/N$. The values of $\alpha(2,2)$ and $\delta_{shuffling}$ show this difference significantly. This is consistent with the conclusion in reference [36] that the growing character of scale-free model is essential to sustain the scale-free state observed in the real systems [36].

In the joint use of MF-DFA and DE, shuffling procedure in DE is essential, especially when we do not know much about the considered spectrum. Un-shuffling DE may leads to serious mistakes. Joint use of shuffling DE and MF-DFA may be a potential tool in the fields of quantum chaos, complex nucleus, and so on, where detecting structure information from spectrum is an essential role.

Regarding the spectrum of a complex network as a time series, we can adopt the powerful tools developed in the field of time series analysis to reveal new features of a complex network.

**Acknowledgements**

This work was supported by the National Science Foundation of China under Grant No. 60274051/F0303 and the Post-Doctor Fund of Nankai University. It was also supported partially by the National Science Foundation of China under Grant No. 10175036. One of the authors (H. J. Yang) would like to thank Prof. Y. Z. Zhuo, Prof. Zhuxia Li and Prof. Xizhen Wu for stimulating discussions.

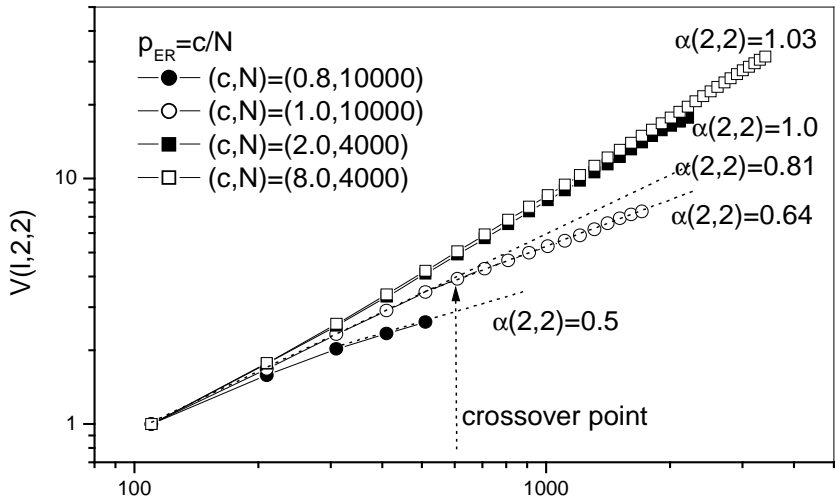

Fig.(1) MF-DFA result for ER model. For $p_{ER}<1/N$, $\alpha(2,2)=0.5$. For $p_{ER}>1/N$, $\alpha(2,2)=1.0$. For $p_{ER}=1/N$, there exist two scaling regimes. The scaling exponents in the short and long regimes are 0.81 and 0.64, respectively. The conventional DFA2 is used.

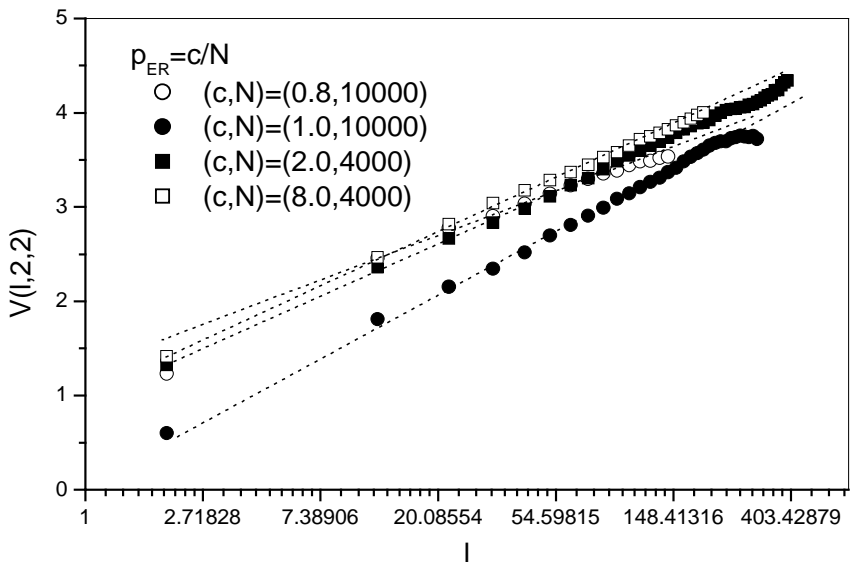

Fig.(2) Shuffling DE result for ER model. For $p_{ER}=0.8/N, 1/N, 2/N, 8/N$ the values of the $\delta$ are 0.44, 0.66, 0.51, 0.54, respectively. At the critical point $1/N$, $\delta>0.5$ significantly. For $p_{ER}>1/N$ or $<1/N$, $\delta\sim 0.5$.

**Table (1)** DE and SDA results for WS small-world model

| Rewiring probability $p$ | Un-shuffling DE $\sigma \pm 0.02$ | SDA $H \pm 0.02$ | Shuffling DE $\sigma \pm 0.02$ |
|---|---|---|---|
| 0.00 | 0.91 | 0.89 | 0.46 |
| 0.01 | 0.87 | 0.89 | 0.55 |
| 0.02 | 0.89 | 0.89 | 0.49 |
| 0.03 | 0.88 | 0.93 | 0.52 |
| 0.04 | 0.86 | 0.88 | 0.54 |
| 0.05 | 0.83 | 0.94 | 0.50 |
| 0.10 | 0.87 | 0.89 | 0.57 |
| 0.20 | 0.84 | 0.86 | 0.54 |
| 0.30 | 0.82 | 0.76 | 0.58 |
| 0.32 | 0.82 | 0.85 | 0.58 |
| 0.35 | 0.84 | 0.85 | 0.53 |
| 0.40 | 0.86 | 0.85 | 0.54 |
| 0.50 | 0.82 | 0.83 | 0.56 |
| 0.60 | 0.84 | 0.81 | 0.57 |
| 0.70 | 0.85 | 0.88 | 0.53 |
| 0.80 | 0.86 | 0.86 | 0.52 |
| 0.90 | 0.88 | 0.87 | 0.57 |
| 1.00 | 0.88 | 0.82 | 0.55 |

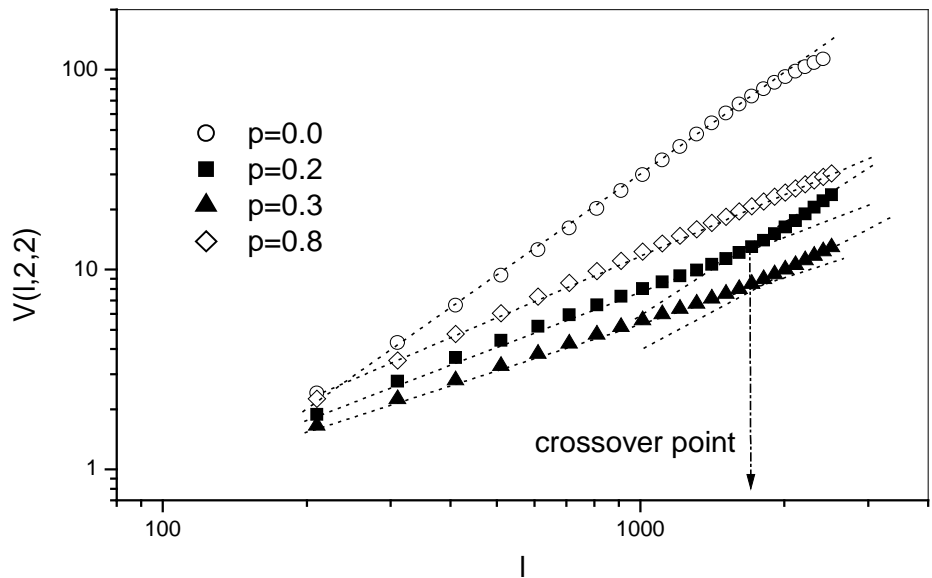

Fig.(3) MF-DFA result for WS small-world model. For p=0.2 and 0.3, there exist two scaling regimes. Conventional DFA2 is used.

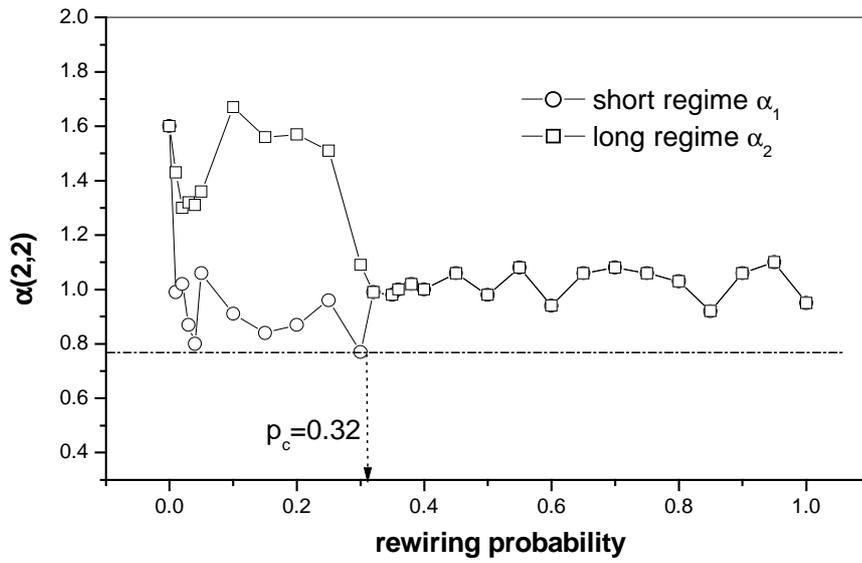

Fig.(4) MF-DFA result for the WS small-world model. Scaling exponents for different p. For 0<p<0.32, there exist two scaling regimes. The conventional DFA2 is used.

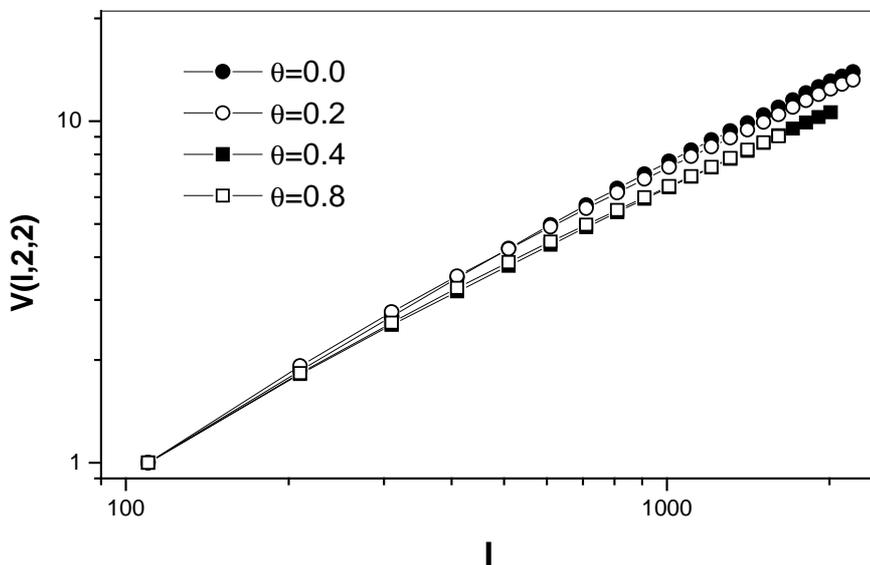

**Fig.(5)** MF-DFA result for GRN networks generated with $\theta=0.0, 0.2, 0.4, 0.8$. A conventional DFA2 is used. A power-law is obeyed almost exactly for large segment size.

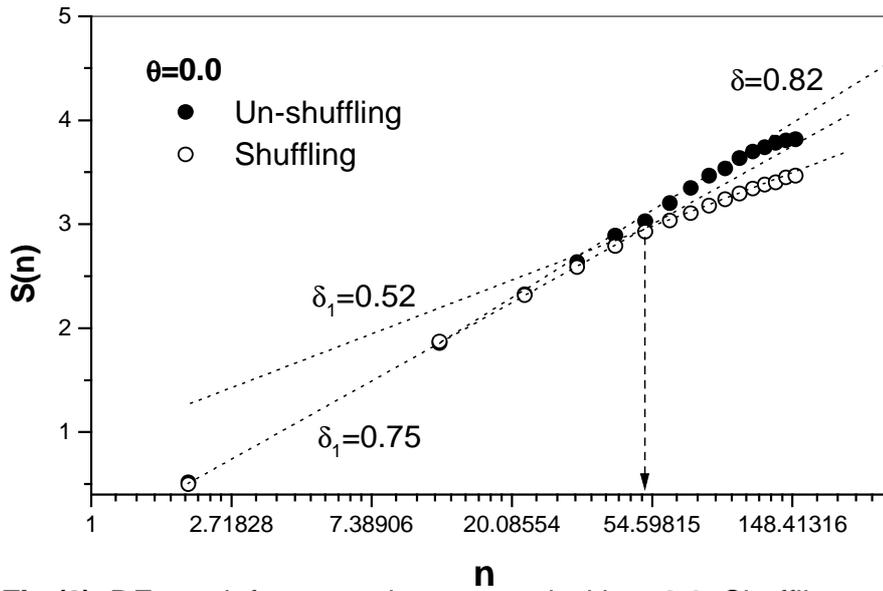

**Fig.(6)** DE result for network generated with θ=0.0. Shuffling result can detect two scaling regions.

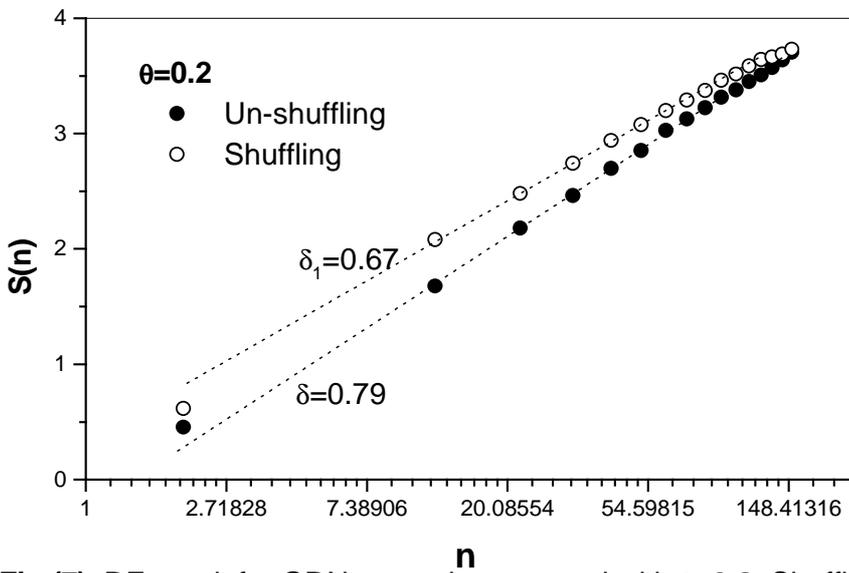

**Fig.(7)** DE result for GRN network generated with θ=0.2. Shuffling result is much smaller than un-shuffling result. A single scaling regime is detected.

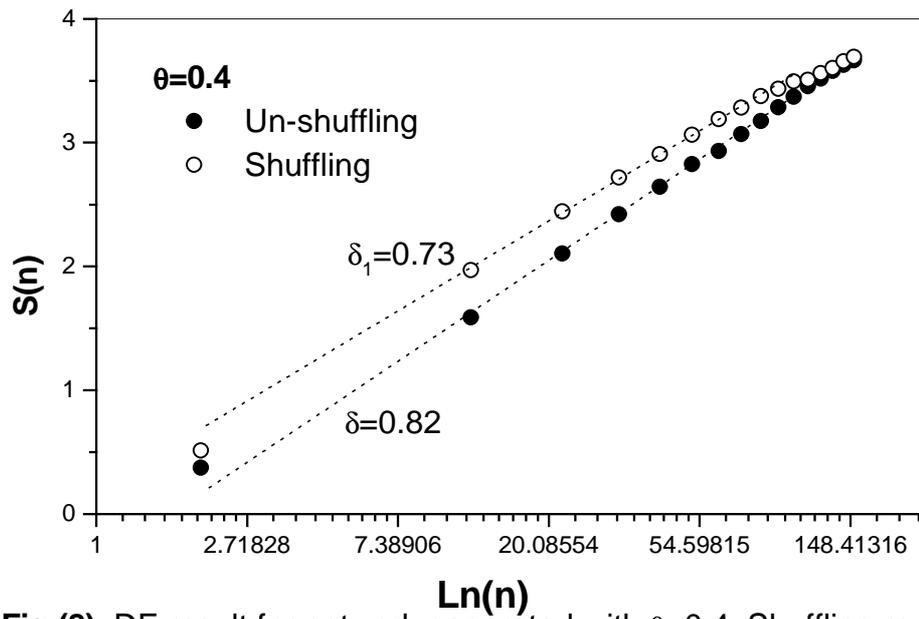

**Fig.(8)** DE result for network generated with θ=0.4. Shuffling result is much smaller than the un-shuffling result. A single scaling regime is detected.

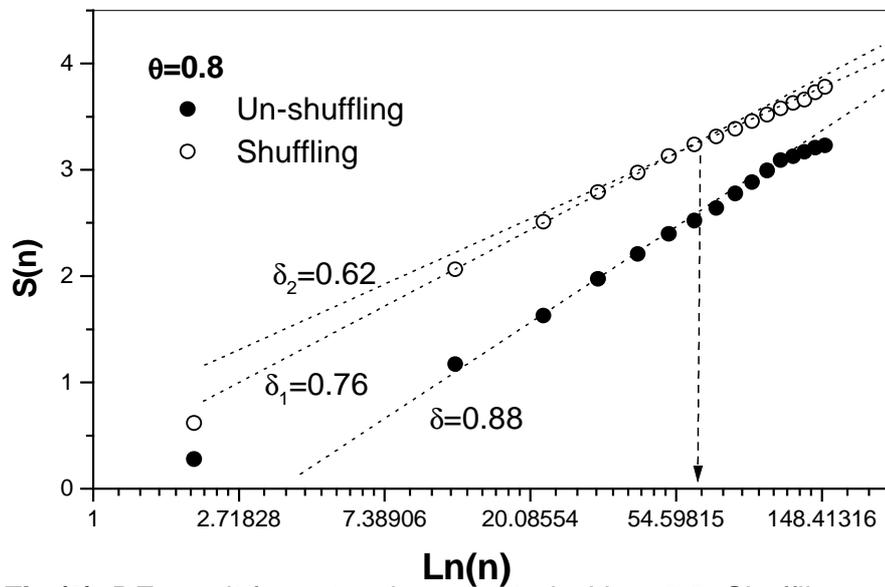

**Fig.(9)** DE result for network generated with θ=0.8. Shuffling result can detect two scaling regions.

**Table (2)** Result for GRN model. Joint use of shuffling DE and MF-DFA to detect PDF form

| $\theta$ | Un-shuffling DE $\sigma \pm 0.02$ | SDA $H \pm 0.02$ | MF-DFA $\alpha \pm 0.02$ | Shuffling DE $\sigma_1 \pm 0.02$ | Shuffling DE $\sigma_2 \pm 0.02$ | Preferred PDF |
|---|---|---|---|---|---|---|
| 0.00 | 0.82 | 0.84 | 0.81 | 0.75 | 0.52 | Levy walk |
| 0.05 | 0.88 | 0.82 | 0.82 | 0.64 | 0.57 | *** |
| 0.10 | 0.80 | 0.82 | 0.78 | 0.74 | --- | Gaussian |
| 0.15 | 0.80 | 0.83 | 0.83 | 0.75 | 0.57 | Levy walk |
| 0.20 | 0.79 | 0.85 | 0.79 | 0.67 | --- | Levy walk |
| 0.25 | 0.80 | 0.85 | 0.80 | 0.75 | 0.51 | Gaussian |
| 0.30 | 0.86 | 0.88 | 0.78 | 0.79 | 0.55 | Gaussian |
| 0.32 | 0.82 | 0.83 | 0.75 | 0.73 | 0.47 | Gaussian |
| 0.33 | 0.69 | 0.84 | 0.77 | 0.75 | 0.61 | Gaussian |
| 0.34 | 0.81 | 0.85 | 0.76 | 0.75 | --- | Gaussian |
| 0.35 | 0.80 | 0.85 | 0.78 | 0.75 | --- | Gaussian |
| 0.36 | 0.74 | 0.86 | 0.77 | 0.75 | --- | Gaussian |
| 0.38 | 0.83 | 0.87 | 0.75 | 0.73 | --- | Gaussian |
| 0.40 | 0.82 | 0.86 | 0.77 | 0.73 | --- | Gaussian |
| 0.42 | 0.83 | 0.88 | 0.76 | 0.82 | 0.58 | *** |
| 0.44 | 0.76 | 0.86 | 0.77 | 0.78 | 0.50 | Gaussian |
| 0.45 | 0.85 | 0.85 | 0.78 | 0.77 | 0.64 | Gaussian |
| 0.46 | 0.82 | 0.85 | 0.78 | 0.77 | 0.67 | Gaussian |
| 0.48 | 0.80 | 0.83 | 0.79 | 0.76 | 0.40 | Gaussian |
| 0.49 | 0.72 | 0.83 | 0.80 | 0.74 | --- | Levy walk |
| 0.50 | 0.87 | 0.83 | 0.80 | 0.77 | 0.52 | Gaussian |
| 0.51 | 0.80 | 0.83 | 0.80 | 0.77 | 0.52 | Gaussian |
| 0.52 | 0.67 | 0.83 | 0.81 | 0.78 | --- | Gaussian |
| 0.54 | 0.84 | 0.86 | 0.78 | 0.78 | 0.55 | Gaussian |
| 0.55 | 0.82 | 0.80 | 0.79 | 0.73 | 0.50 | Levy walk |
| 0.60 | 0.85 | 0.85 | 0.78 | 0.70 | 0.53 | Levy walk |
| 0.65 | 0.87 | 0.88 | 0.76 | 0.78 | --- | Gaussian |
| 0.70 | 0.90 | 0.82 | 0.80 | 0.77 | --- | Gaussian |
| 0.75 | 0.95 | 0.85 | 0.79 | 0.75 | 0.55 | Gaussian |
| 0.80 | 0.88 | 0.83 | 0.76 | 0.77 | 0.62 | Gaussian |
| 0.85 | 0.88 | 0.87 | 0.75 | 0.73 | --- | Gaussian |
| 0.90 | 0.75 | 0.85 | 0.74 | 0.82 | --- | *** |
| 0.95 | 0.86 | 0.88 | 0.80 | 0.79 | 0.42 | Gaussian |
| 1.00 | 0.81 | 0.88 | 0.76 | 0.79 | --- | Gaussian |